\documentclass[A4,11pt]{article}

\usepackage[T1]{fontenc}
\usepackage{authblk}
\usepackage[numbers,sort&compress]{natbib}
\usepackage{natbib}

\usepackage{times}
\usepackage[left=1in, right=1in, top=1in, bottom=1in]{geometry}

\usepackage[utf8x]{inputenc}

\usepackage[dvipsnames,table]{xcolor}

\usepackage{hyperref}

\hypersetup{%
  colorlinks=true,
  linkcolor=black, 
  urlcolor=blue, 
  citecolor=black, 
}
\usepackage{lastpage,fancyhdr,graphicx}
\usepackage{ifthen}
\newboolean{isdraft}
\setboolean{isdraft}{false}

\ifthenelse{\boolean{isdraft}}{%
  \newcommand{\new}[1]{{\color{blue}#1}} 
}{
  \newcommand{\new}[1]{#1}
}

\begin{document}

\title{Ten Simple Rules for Attending Your First Conference}

\author[1]{Elizabeth Leininger}
\author[2]{Kelly Shaw}
\author[3]{Niema Moshiri}
\author[4]{Kelly Neiles}
\author[5]{Getiria Onsongo}
\author[6]{Anna Ritz\thanks{Corresponding Author: aritz@reed.edu}}

\affil[1]{Division of Natural Sciences, New College of Florida, USA}
\affil[2]{Computer Science Department, Williams College, USA}
\affil[3]{Computer Science \& Engineering Department, University of California San Diego, USA}
\affil[4]{Department of Chemistry \& Biochemistry, St. Mary's College of Maryland, USA}
\affil[5]{Mathematics, Statistics \& Computer Science Department, Macalester College, USA}
\affil[6]{Biology Department, Reed College, USA}
\date{\today}

\maketitle

\section*{Introduction}

Conferences are a mainstay of most scientific disciplines, where scientists of all career stages come together to share cutting-edge ideas and approaches. If you do research, chances are you will attend one or more of these meetings in your career. Conferences are a microcosm of their discipline, and while conferences offer different perspectives in different disciplines, they all offer experiences that range from a casual chat waiting in line for coffee to watching someone present their groundbreaking, hot-off-the-press research. \new{Here, we share tips for trainees and their mentors. Our recommendations are based on our experiences of attending conferences and mentoring students to improve their conference experiences.} As you head to your first scientific conference, these rules will help you navigate the conference environment and make the most of your experience. 

\new{\paragraph{In-person Conferences:} Scientific meetings have historically been in person, where all attendees travel to one location to share and learn about recent work in the field. In-person conferences provide extensive opportunities to meet and interact with other researchers. Our rules include considerations for all parts of attending an in-person meeting, including deciding whether to attend and present research, navigating travel and the conference schedule, and debriefing after the conference.}

\paragraph{Virtual Conferences:} During the COVID-19 pandemic, many conferences \new{have been and continue to be} offered remotely~\cite{olena2020covid,sarabipour2020research,sarabipour2021changing}. Even before 2020, some conferences had already moved to a virtual format in response to climate change~\cite{ligozat2020ten} and/or travel bans~\cite{reardon2017latest}, \new{noting the environmental, financial, and geopolitical challenges posed by in-person meetings. Virtual conferences are more inclusive than their in-person counterparts in many ways: more people can attend them from all over the world, virtual meetings are less costly and easier to organize, and attendees can participate from the comfort of their homes~\cite{sarabipour2021changing,salomon2020future,rich2020ten,arnal2020ten}. Given that virtual conferences are becoming more }commonplace~\cite{gichora2010ten,lortie2020online,sarabipour2021changing,rich2020ten,arnal2020ten}, we include considerations for each rule when attending one.

\paragraph{For Mentors:} \new{Academic mentorship takes many forms, including advising student research, teaching students in courses, and serving as a department chair or organizer for student activities. In many fields, graduate students mentored by a principal investigator (PI) are expected to disseminate their research at scientific meetings.} Conference attendance is a valuable experience for undergraduate students, whether they are presenting research~\cite{mabrouk2009survey} or just beginning to learn about the field~\cite{davis2017supporting,wright2019can}. As mentors, you might be working with undergraduates, graduate students, or other staff \new{who may be} new to conferences. We offer guidance on ways for you to help mentees navigate the subtleties and assumptions of your field. 

We have also developed a web portal (\nameref{webportal}) which contains far more information about these rules, tables of professional societies and conferences in different disciplines, and other resources that may come in handy for first-time conference attendees and their mentors. We encourage any reader to use, adapt, and contribute to these materials.

\section*{Rule \#1: \new{Select a conference that aligns with your goals}}

Why go to a conference, anyway? When you are deciding whether to attend a conference, consider the experience and how it will help you develop intellectually and professionally. There may be more than one answer! Conferences are a ``meeting of minds'' -- a place for researchers to gather, present their research, give feedback on others' research, engage in professional development, and network with one another. 

Conferences differ greatly in size and scope. While some conferences provide a large overview of an entire academic discipline with tens of thousands of attendees, other conferences focus on specific subdisciplines in an informal and personalized setting. \new{Smaller conferences are excellent venues for networking, sharing your data informally, etc.} Some conferences are national or international in scope, while others have a regional focus. All of these variables impact factors such as the conference location, scope of presented research, the cost of attendance, and more. 

If you are planning to attend a conference, you might also plan to present research. You \new{and the others who have worked on the research would} make this decision before the conference, often many months prior, and you typically submit materials to be considered for presenting. Consult with your mentor and colleagues in your research group about the benefits of and opportunities for presenting your work. \new{In some fields, such as biology or chemistry, you might submit a short abstract for a poster presentation or a talk. In other fields, like computer science, you might submit a full paper that is peer-reviewed and published upon acceptance.} Conventions vary by discipline; make sure to discuss norms in your field with your research mentor \new{and all other authors of the work, who need to approve any submission.} 

In addition to contributed talks and posters, conferences provide formal and informal opportunities for professional development and networking (See Rule \#7). Review the professional development opportunities at the conference you are attending, and participate in opportunities that sound interesting and align with your goals. 

\paragraph{Virtual Conferences:} \new{Nearly every annual meeting in 2020 and 2021 has shifted to a virtual format due to the COVID-19 pandemic. Many traditionally in-person conferences aim to retain multiple goals of their conferences -- a venue for presenting the latest work, networking, and professional development -- in a virtual format. Alongside these legacy meetings, the virtual format has also spurred new grassroots virtual conferences~\cite{rich2020ten}.} For example, conferences such as Neuromatch and Black in Neuro are newer additions to Neuroscience conference offerings that originated as virtual experiences. \new{Some virtual conferences offer more focused research areas than the larger meetings, since they are generally easier to organize and manage. They also have fewer barriers for attendance (allowing researchers from all over the world to attend), they are less wasteful (in terms of money and environmental impact), and they are overall more equitable than in-person conferences~\cite{sarabipour2020research,sarabipour2021changing}.  Consider all of these aspects when deciding on a conference to attend.}

\paragraph{For Mentors:} Empower students to attend conferences. Suggest conferences for your students that match their present goals/professional development needs \new{and explain how they could benefit from attending these meetings.} Be aware of socioeconomic privilege and implicit bias when recommending opportunities to trainees (e.g., recommend equally and equitably, help them find funding, and connect them to your professional network at conferences). 

\section*{Rule \#2: Find others to foot the bill}

\new{Organizing and running conferences cost money, and attendees usually pay to attend these events. While registration costs are often low for virtual conferences, in-person conferences can get expensive quickly due to additional costs associated with travel, room and food expenses, poster printing fees, childcare, appropriate clothing, etc. Therefore, it's important to have a funding strategy to attend an in-person conference. Funding sources are often available to offset some (or even all) of your costs, and knowing where they exist early in the process will help you take advantage of these opportunities.}

\new{First,} some conferences offer discounted registration or travel grants for students, attendees from countries with traditionally fewer economic resources, etc. \new{Some travel grants may include additional support for trainees with disabilities or trainees with small children. Other societies are establishing separate child care grants for trainees with families.} Often these funding opportunities are made possible by professional societies, grants from organizations such as the National Science Foundation, or by sponsorship funds provided by industrial partners. An in-person conference might have volunteer opportunities where you are compensated with a waived registration fee or housing for working at the conference. \new{Second, many academic institutions have funds intended to support student travel.} Principal investigators may have a budget to support their students' attendance at a conference. Research scholarships and/or travel awards may be available at the departmental or college level. Sometimes these opportunities are not widely advertised so it is always good to ask around. If you do not have a research mentor, asking the department chair may be appropriate to help you navigate your search. Another place to look is student clubs and organizations, which may have funding to support students presenting research in a field relevant to the group's goals. \new{Third,} national societies, often the same societies that are organizing the conference, may provide support for student travel at the society, division, and regional chapter level. Some of these opportunities are dependent on student demographics such as being first-generation, a student from a historically underrepresented population, or a woman in the field. \new{Information about these awards is generally not in a central location, so visit each level's website to search for potential sources (the web portal includes a list of societies in STEM, many of which provide travel funds).}

An important note: many of these awards and funding opportunities \new{for in-person conferences} will often require that you front the money and then get reimbursed after the conference. This means you would need to put things like airfare, hotel reservations, and the like on your own credit card and then submit receipts for reimbursement.  If this presents a hardship for you, reach out to your department for help. This is not the first time, nor will it be the last, that your department has encountered this problem so you should not hesitate to ask for assistance in navigating this process. Keep track of all itemized receipts for the conference, and ask the funding provider for details about reimbursement.

\paragraph{Virtual Conferences:} \new{Virtual conferences} generally cost less than their in-person counterparts~\cite{sarabipour2021changing}, and many virtual conferences have \new{reduced or} waived the student fees associated with conference registration. If paying for a virtual conference registration is a hardship, there is no harm in asking the conference organizer if there is a way to attend. 

\paragraph{For Mentors:} Funding support deadlines for conferences are often so far in advance that students, especially undergraduate students, \new{may miss these opportunities. Help students identify and meet these deadlines.} If your institution has funding sources for student travel but those sources are only available to those students who ask (or whose mentor's ask), it might be time to suggest a conversation about equity of those opportunities. This policy may disproportionately disenfranchise students from historically underrepresented populations who are not as well versed in the hidden curriculum of academia and thus do not know to ask for these opportunities~\cite{smith2013mentoring}. 

\section*{Rule \#3: Know your logistics}

\new{It is important to be aware of the \textit{Who}, \textit{What}, \textit{When}, \textit{Where} logistics (see Rule \#1 for \textit{Why}) to ensure a smooth conference-going experience. Such logistics will vary depending on if you choose to attend an in-person conference or a virtual conference, or even depending on the format of virtual conference you choose to attend.}

\textit{Who} do you know will be at the conference? Housing and transportation costs for conferences can be expensive, complicated, and/or both. Consider traveling together with people you know and sharing housing costs by rooming with labmates, acquaintances, etc. Some conferences and field-specific societies offer roommate-matching services for those seeking conference roommates. You may also be traveling with family or dependents; check to see if the meeting has programming or resources for companions, childcare and lactation facilities, etc. as needed.

\textit{What} will you need to bring to the conference? It is usually preferable to pack light, and pack for the situations and weather you will encounter (see Rule \#4). Double-check baggage policies for your mode of transportation, as policies vary drastically by carrier and fare. If you are presenting, make sure that your presentation materials travel safely with you. If you are giving a talk, make a few different backups of talk slides and media (store on a thumbdrive or in the cloud) in case your computer is lost, stolen, or broken. \new{Posters can be transported in a poster tube or even printed on fabric to fold in your luggage. The poster can also be shipped to the venue or printed on-site if it is challenging to travel with it.}

\textit{When} do you need to make key conference travel decisions? Chronologically, these include: writing and submitting a research abstract or paper, having your abstract or paper accepted, registering for the conference, making travel and housing arrangements, and actually traveling to the conference and presenting your work. Because abstract and paper submission deadlines can fall several months ahead of the meeting date, it is important to be aware of abstract deadlines for conferences in your field. When scheduling travel, make sure that you arrive before, and leave after, your scheduled presentation date/time. \new{If your travel has multiple legs of transportation, give yourself enough time to make each connection.} International travel may require visas, and the lead time \new{and expense} varies greatly depending on the origin and destination countries. \new{Consider visas and passports in both your timeline and your budget, and talk with your mentor about whether an international conference is feasible.} Conferences with an international scope will usually communicate information about visas and lead time. 

\textit{Where} is the conference, and where will you stay while attending? Depending on the scope of the conference (see Rule \#1), housing may be easily obtainable at the conference venue, or a mad scramble between all 30,000 attendees to find a room near the convention center. Conferences will typically contract guaranteed rates at nearby hotels; heed deadlines and sign up early if you possibly can. \new{Be aware of ADA accessibility when booking if you require rooms with accessible features.} Acquaint yourself with the transportation options to and within the conference host city. Again, conferences will usually send information and tips regarding transportation, housing, and the like. 

\paragraph{Virtual Conferences:} 

\new{The \textit{Who}, \textit{What}, \textit{When}, \textit{Where} logistics of virtual conferences are also extremely helpful to determine before the meeting. }

\new{Identify \textit{Who} you know at the meeting, and make plans to meet with them virtually during a break. Virtual conference platforms allow you to browse the list of presenters and/or conference attendees (see Rule \#7). Know \textit{What} you are expected to do in different sessions; if you plan to attend a virtual panel, for example, think about some questions you might want to ask the panelists. }

\new{Familiarize yourself with \textit{When} the conference activities are, and if any action is required for accessing the conference materials. The format of virtual conferences often differs from the time-limited experience of in-person conferences. For example, they may run longer to allow for a different pace of engagement, and they may have components that are synchronous (e.g. talks scheduled live for specific times) or asynchronous (pre-recorded talks for on-demand viewing). Some virtual conferences are offered completely asynchronously, making them more inclusive for attendees with care responsibilities or in time zones across the world~\cite{arnal2020ten}. Make sure to note conference dates and time zones of scheduled synchronous sessions as appropriate. If presenting at a virtual conference, pay very close attention to announcements regarding presentation logistics; you may need to upload your presentation before the formal start of the conference and check in at various times to field questions. }

\new{Finally, be organized by knowing \textit{Where} the conference activities are happening, often through a web interface or a collection of Zoom links. If you are unsure about the digital platform or need extra assistance, contact the organizer who will be able to help you get help before the conference. If the virtual conference requires high internet bandwidth, make sure your setup at home allows for participation. }

\paragraph{For Mentors:}  If you are also attending the meeting, make sure that your students are at the same point logistically as you (register for the conference together, book flights together, etc.) If you are funding the students, putting as much of the costs on your business card up-front reduces their financial burden. If you are not attending the meeting, reach out to your colleagues who will be there to help arrange housing and other connections for your students.

\section*{Rule \#4: Prepare for the environment}

A good rule of thumb is to dress as if you might meet your future employer at a conference (which you very well might do!). At the same time, wear something you are comfortable in. Different meetings and disciplines may have a variety of dress code standards. Attendees who are presenting may wear more formal clothes, but many people will dress as they do when they are at their home institutions.

\new{When you are packing for an in-person conference, check the weather before you go, since you may be outdoors (e.g., to reach the conference from the airport, to walk to find dinner, or participate in planned conference outings). Between moving around the conference venue, seeking out meals, and milling about poster sessions, you will spend a lot of time being active. Wear shoes and clothing that allow you to comfortably do all these things. If you have any questions about ADA accessibility at the conference, contact the conference venue (e.g., if it is a conference hall attached to a hotel) or contact the conference organizer (the conference website will contain this information). Some conferences have an app (See Rule \#6) which may have a Q\&A Forum. Finally, conference days can be long - bring a water bottle to stay hydrated, a light snack to sustain you between meals, and a comfortable bag to keep belongings.}

\paragraph{Virtual Conferences:} You will likely be in the comfort of your own home during the conference. Realize it is easy to sit and stare at the screen for hours on end. Be sure to take breaks, stay hydrated, stretch, and focus your eyes away from your screen every so often. It is also a good idea to test your video, audio and background before the conference. Find a room with good lighting and make sure your background is not distracting. Some video conferencing platforms have video settings that will enhance your image if the lighting is low - try them out. \new{Conference days can be extremely long in virtual platforms -- sometimes longer than typical in-person schedules -- so be sure to pace yourself (see Rule \#6).}

\paragraph{For Mentors:} Talk to your students about the level of formality at the meeting. If you know that \new{an in-person conference} will have outdoor events, mention that. If you are comfortable with sharing, tell your students the types of things you pack for conferences. 

\section*{Rule \#5: Learn how to take in the science}

Science communication happens in a variety of ways at a conference. There are almost always formal and informal mechanisms for learning about and presenting science. Even within formal mechanisms, there is a range of how attendees communicate their research. Some speakers are invited to give \textit{keynote presentations}, which often provide a perspective of the field or cover a broad range of projects. Other speakers are selected to give \textit{technical presentations}, based on submitted abstracts or full papers. Both keynotes and technical presentations provide time for asking questions at the end, and attendees may line up in front of a microphone or raise their hand depending on the number of people. Finally, \textit{poster presentations} offer a more interactive way to talk with researchers, where attendees walk up and chat with authors standing in front of their posters. 

But wait, there's (often) more! Many conferences offer tutorials, workshops, or special sessions that are held either right before or right after the main conference, usually at the same location. These topics may be even more narrowly focused than those at the conference, or involve emerging concepts in the discipline. There also may be forums run by graduate/undergraduate student societies, designed for students. Note, though, some of these may require registration beyond the main conference.

So, how do you actually ``take in the science?'' \new{First, know that it is often impossible to see everything - the conference may have multiple conflicting sessions, the days are quite long, and you need to take time to rest and recharge.} As you make a conference plan \new{to prioritize what you want to see} (see Rule \#6), let the conference program be your guide. Find a way that works for you to take notes on what you learn throughout the conference. We, the authors, have developed note-taking strategies that involve carrying around a notebook, using Google Docs, writing in the conference program, taking notes on our smartphones, making sketchnotes, even sending ourselves emails. In addition to the science, take notes about whom you meet - you never know if you will meet them again or if you might want to connect with them after the conference. Find something that works for you. \new{Some conferences may have rules about taking digital photos or videos of talks and posters, so be sure to check beforehand if you want to do either.}

\new{At in-person conferences,} there are a plethora of informal ways to interact with other researchers and learn about their science. Researchers appear to have a common need for caffeine, and coffee breaks are a staple of every conference. If you are not a coffee drinker, there are often other warm beverages and snacks on hand. Conferences may also offer breakfast or lunch, and each one is in an invitation to meet someone new (see Rule \#7). Many conferences have a ``big-ticket'' event - an evening banquet, an outing to a tourist attraction, a social hosted by a sponsor, etc. If the conference has a big-ticket event, it is not one to miss! There may also be a career fair, special meetings for first-timers, or meet-ups for special interest or affinity groups. Take advantage of these to find other attendees with similar interests, backgrounds, and experiences (see Rule \#8).

\paragraph{Virtual Conferences:} \new{Virtual conferences have the enormous benefit in that the talks are almost always recorded (even when they are presented synchronously to all attendees). Further, sometimes conferences will rebroadcast talks 12 hours later for attendees in different time zones. The material from conferences often remains on the virtual platform after the conference concludes, allowing attendees to watch and rewatch talks at their leisure.} In addition to the formal presentations, \new{virtual} conferences have developed strategies for the interactive conference portions (e.g., poster presentations, breaks, and social events). \new{Some digital platforms offer mechanisms to ask questions and continue discussions after a talk has concluded. Cold Spring Harbor meetings, for example, have a Slack workspace with channels for each session to continue Q\&A and networking.  The chat functionality also allows researchers to talk to poster presenters throughout the entire meeting and send short messages about appreciating people's work in a way that cannot happen at an in-person conference. Take advantage of the interactive parts of the conference that can help support your learning and professional development.}

\paragraph{For Mentors:}  Share the events you plan to attend with your students, and give advice about what talks or events will be most useful for them. \new{Encourage your students to ask questions, either in the Q\&A after a talk or less formally at the poster session. } If you have students or colleagues who are presenting, encourage your students to attend those talks. 

\section*{Rule \#6: Make a conference strategy}

The idea of a conference - days of uninterrupted learning about fascinating ideas and exchanging insights with other folks excited about the topics you are passionate about - \new{might sound} like a dream come true! The reality is that conferences can be exhausting if you do not have a plan for selectively attending activities that will provide you the most benefit and for practicing self-care. \new{This is especially important for neurodivergent attendees who may get overwhelmed by the amount of information conveyed during a conference.} It's good to fashion a draft plan several weeks before the conference in which you prioritize the events you want to attend. Having a written (or app-built) plan with scheduled breaks gives you a solid framework that you can tweak on the fly as new opportunities appear. If you intend to participate in pre- or post-conference events, it is important to factor in extra self-care to compensate for the increasing duration.

Before making a conference strategy, gather resources to help you make a plan. Conferences usually have a website with lots of information, and often release a schedule overview as the keynote and technical talks and posters are finalized. The conference may make use of an app, where attendees can find the schedule of events and connect with others, \new{and/or you may receive a conference program.} These are all great resources to have on hand as you make your strategy, \new{and it is useful to skim the conference program before the meeting begins.}

How do you prioritize what to attend? First, it is good to attend keynote and panel sessions as they provide perspective into the wider concerns of your field and often are forward looking to emerging challenges. Second, definitely attend technical presentations related to your specific area of focus in order to know what research is being done and become part of that community of researchers. Reading papers or watching videos in advance and thinking of what questions you might like to ask about the work is a great way to prepare so that you can contribute to the discussion in a positive way. \new{Third, the poster sessions are often short, so make sure you know which posters you want to visit while the presenter is there.} Fourth, if the conference offers any first-time or new attendee events, plan on attending those as you will make some connections with other attendees that will make the conference more enjoyable and less lonely. Finally, attending the networking events (see Rule \#7) \new{helps you get} to know your colleagues as individuals on a personal level (not all discussions are about the research) and also exchanging your research ideas. 

Working self-care into your plan is essential - but how? First, if you have a daily ritual such as exercising or going for a walk, sustain that ritual during conference days \new{and decide when you will fit this activity in. If you are at an in-person conference,} find out if the hotel has the equipment you need for exercising; many downtown gyms or community centers sell day passes. Second, recognize that your brain is going to need breaks between talks. \new{If the conference is not in your native language, your brain may be working overtime to process the science.} Determine time slots when the presentations are not of particular interest to you and plan on taking a walk, getting a coffee, or doing some other activity that will help you recharge.  You will see lots of experienced conference attendees disappear for stretches of time for exactly this purpose. For many of us, traveling to \new{in-person} conferences is a way to experience new parts of the country or the world. It is fine to take a short break to experience \new{local destinations} such as a museum, park, cultural attraction, etc. 

\paragraph{Virtual Conferences:} When attending virtually, it may be hard to devote entire days to attending the conference. Prioritize attending the synchronous events such as the keynotes \new{and} panel discussions, \new{especially if they are not recorded for later viewing.} Take advantage of videos posted for areas you are interested in and watch those ahead of time so you can choose to attend the sessions where you're interested in the question and answer portion.

\paragraph{For Mentors:} Encourage your students to come up with a written plan and discuss it with them before the conference. Suggest specific presentations that would be good for them to attend and explain to them the importance of strategic downtime and self-care.

\section*{Rule \#7: Make new friends but keep the old; be ready to communicate}

Conferences offer a great opportunity to exchange ideas, network, and potentially form collaborations with other researchers. Networking opportunities can be organized or spontaneous. Organized networking events may include events such as socials, affinity group meetings, or mentorship opportunities that pair newcomers with established researchers in the field. Many researchers also make efforts to meet with current or prospective collaborators at meetings. In addition to organized events, lots of networking happens spontaneously -- waiting in line for coffee or tea, at a poster presentation, etc. Take advantage of the opportunity to meet lots of people who are interested in sharing their science and forming new professional connections. 

Pre-plan a few people you would like to meet at the conference. A good way to find people is to look at the \new{conference program, which contains presenter names and sometimes includes a full list of attendees }(See Rule \#6). Start with the research area you are interested in. Typically, the first author presents the work and will most likely be at the conference. Read their work before the conferences and prepare questions to start a conversation. If they have a research website, they will likely have a picture. Find out what they look like in advance. \new{In-person} conference attendees usually wear name badges which helps if you're trying to meet someone whom you've never met in person. \new{If they are giving a talk, try to attend; following up with the speaker after their talk is a good way to strike up a conversation with them.} If you did not get a chance to talk with everyone you wanted to, you can always follow up with them over email (see Rule \#10). You can use the conference to make it not feel like a cold email - tell them you attended their talk, saw their poster, etc., and ask any questions. Researchers delight in being asked informed and insightful questions about their work.

\new{In-person conferences have many opportunities for informal networking, but it still may feel awkward to strike up a conversation.} If you are coming from a lab or institution with colleagues also attending the conference, agree to introduce one another to whomever you are speaking with. Also, most conferences have events just before the official program starts such as breakfast or socializing events in the evening. These are a great way to meet people in a more relaxed environment. If you are feeling overwhelmed or out of place, which is normal (see Rule \#8), it is okay to spend time with people you already know.

\paragraph{Virtual Conferences:} Introducing yourself and having a one-on-one conversation \new{can also be} challenging in virtual environments. \new{ Networking still happens at virtual conferences, but the interactions are necessarily more intentional than striking up a conversation during a coffee break. Many virtual conference platforms offer ways to interact, and conferences organize social events to foster this networking.} If \new{the online conference } platform has social events or research discussion groups, join in. Leave your camera on, if comfortable, and participate in the chat, if appropriate. Some virtual conference platforms have messaging capabilities, which you can use to have conversations with individuals or small groups. 

\paragraph{For Mentors:}  If you will also be attending an \new{in-person} conference, be intentional about introducing your students to your professional network. Invite them to join a group meal you plan on attending with people you already know and introduce them to colleagues. If you are talking to someone working in an area your student is interested in, and you see your student close by, be sure to introduce them. \new{Pointing students towards posters that are relevant and interesting to them is also a great way to help attendees begin conversations with researchers.}

\section*{Rule \#8: \new{Prepare to (safely) get out of your comfort zone }}

For most of us, meeting new people or joining a new community can be nerve-wracking or intimidating, even when you have a lot in common with people in the community. Things \new{can be even more challenging} if you come from a historically underrepresented community and do not see yourself represented at the conference. Academia, and by extension its conferences, is a traditional and elite institution whose diversity, or lack thereof, can make it a less than welcoming place at times. Imbalance in representation is gradually being acknowledged by assessing the demographics of invited speakers and society awardees~\cite{sarabipour2020research,martin2014ten,le2020analysis,shishkova2017gender}. If you are feeling out of place, know you are not alone, and that it can take some time to feel comfortable. Try to meet new people and make new friends. You will likely see the same people if you attend the conference again. And, the more conferences you attend, the easier it gets. Some conferences host affinity group events, which is a good opportunity to meet and network with other attendees in a safe and welcoming space. Our web portal hosts a crowd-sourced list of affinity groups. 

Come up with a plan for how to talk about your interests and research. Have a 90 second elevator pitch prepared for your initial introduction and be prepared to switch into a slightly extended version of that pitch if the other person expresses interest. Because it is important to learn about the other person (people love good listeners), make sure to ask them about their research, including asking follow-up questions after their initial answers to show your interest. If the exchange goes well, create a note of their names and affiliations for use later. Also, if you see them later in the conference, acknowledge their existence with a smile or head nod.

\paragraph{Virtual Conferences:} An advantage of virtual conferences is you will be in your own space. Take advantage of this opportunity in a virtual setting to make yourself comfortable (See Rule \#4). It is okay to turn off your video if you need to step away for a few minutes. 

\paragraph{For Mentors:} It can be helpful for mentees to know they are not alone in feeling out of place. Normalize the experience of feeling nervous and out of place. Let them know conferences can be awkward and that is okay. If you are mentoring students from underrepresented groups, being culturally competent will help you better support them. \new{Cultural competency is the ability to work effectively with people from different cultural backgrounds.} In examining the role of cultural competence in a biology classroom, Tanner and Allen highlight the importance of cultural competence in creating an inclusive and welcoming environment for students from underrepresented groups~\cite{tanner2007cultural}. A good place to start is finding out if your institution offers cultural competency training. If it does not, the Association of American Colleges \& Universities offers cultural competency resources~\cite{AACU}. 

Finally, mentor your students -- particularly marginalized students who may find academia and conferences less-than-welcoming -- through the conference experience and their overall professional development with ``compassion, advocacy, and support''~\cite{singleton2020open}.\new{ Be aware that imposter syndrome stems from systematic bias, and calling out imposter syndrome sometimes can have the opposite effect for mentees~\cite{tulshyan2021stop}}. Become involved in helping meetings achieve gender and racial balance in their selection of awardees and keynote speaker invitations~\cite{martin2014ten}. See our web portal for more resources on cultural competency and improving representation at conferences.

\section*{Rule \#9: \new{Take charge of your social interactions}}

Conferences bring together many people from all over the world, and navigating a complex professional-yet-social environment can be challenging. All members of a scientific community have a responsibility to help make a welcoming environment and should in turn feel welcomed. As a conference attendee, you are a member of the scientific community. \new{At the same time, we acknowledge that power imbalances may be prevalent at the conference, often reflective of career stage~\cite{jackson2019smiling}}.  Many societies have established codes of conduct to which attendees must adhere~\cite{favaro2016your}, and the conference should include a contact if you witness or are subjected to troubling behavior. Above all, you should never feel unsafe or pressured to participate by anyone. 

\new{Before you travel to in-person conferences, acquaint yourself with the code of conduct and have a contact person (e.g., your mentor, someone from your institution, or a friend) that you can reach out to. You may need to prepare yourself to adjust to the cultural norms of the conference location, if it is notably different from your academic environment. While there are social opportunities at in-person conferences, remember that these are professional events.}

\paragraph{Virtual Conferences:} Social situations are different in a virtual platform. There may be fewer awkward conversations, but online interaction can pose its own problems. For example, you may see unwelcome visitors hack into presentations or observe uncourteous behavior from other attendees. You may also find some features (like the chat function or unmuted attendees) distracting during a presentation. When participating in virtual conferences, follow any instructions session chairs give about how to interact during presentations. As a general rule, make sure you follow online etiquette (such as muting yourself on Zoom unless speaking). \new{Attendees should observe the same codes of conduct in a virtual setting that they would observe in an in-person conference.}

\paragraph{For Mentors:} It is important that students have your contact information (or the information of someone trusted who will be at the meeting). Stay up to date on your organization's code of conduct. If your society has not yet established a code of conduct (check on our web portal), encourage the leaders to do so and support their efforts~\cite{favaro2016your}. 

\section*{Rule \#10: Tie up the loose ends after the conference}

After attending the conference you will likely want to come home and collapse from all of the excitement, but wait\ldots you have a few more things to do before you are done. \new{These loose ends pertain to both in-person and virtual conferences.} First and foremost, update any notes you have about the conference itself before you forget. Conferences can be a fast-paced blur, so make sure to record any feedback you get on your work at the conference, so that you have it when you are ready to make improvements. Updating your resume and/or curriculum vitae (CV) is another important step you should complete as quickly after the conference as possible. In addition to any posters or presentations you gave, you should also add any awards you received including travel funding awards. Depending on your career stage, you might also include other events or support you received - check with a mentor about what is appropriate. 

Next you should follow up with people you met at the conference. This helps to solidify the relationships you began at the conference. Email people you are interested in speaking with again and ask for an opportunity to meet. If you spoke with companies or potential job seekers, follow up with an email containing your resume and statement of interest. When you attend a conference you meet so many people that it is hard to remember everyone. By reaching out with a simple note or a LinkedIn invitation, you will be helping people to remember you which can lead to future collaborations and/or job opportunities!

If you received a travel award \new{(for an in-person conference, for example)} that requires reimbursement you will need to carefully follow the instructions and/or rules of your funding agency. Reimbursements can take quite a while to process so the faster you get it done after the conference, the better. Pay special attention to rules regarding the submission of receipts (itemized receipts are required in some cases) and deadlines for submission (some agencies require reimbursement documents be submitted within a month of the conference ending). \new{Make a note to yourself to follow up on reimbursement; if} you do not hear back about your reimbursement within two weeks, go ahead and reach out to inquire about your paperwork (you can also ask your mentor if it has been longer than you were expecting). 

Finally, you should take time to write thank you notes or emails to anyone who supported your travel to the conference whether that be financially, in conference preparation, or in your research. This not only helps you to further strengthen relationships with those people, but also helps future students receive these awards by leaving a good impression with the awarders. 

\paragraph{For Mentors:}  \new{Communicate your institution's rules and requirements about reimbursements to your trainees, and tell your trainees who to contact for reimbursement paperwork (e.g. a department administrator).} We suggest holding a paperwork session after the conference so that all travelers can come together and fill out paperwork, print and/or make copies of necessary materials, and ask questions of each other and you. If multiple travelers attended the same conference, then mentors can also submit all of this paperwork together to increase the chances of its timely completion. We sometimes also bring a giant box of thank you notes so that students can easily grab one to write a note of thanks to those who helped them along the way. If students run into bureaucratic difficulties with paperwork, help them get the issue resolved.

\section*{Conclusion}

Scientific conferences are an amazing intellectual and professional opportunity, though attending one for the first time may be overwhelming. We hope that first-time conference attendees will feel more empowered and prepared to get the most out of their experience\new{, be it virtual or in-person,} with knowledge of these ``unwritten rules.'' Happy travels!

\section*{Supporting information}

\paragraph*{SI Web Portal}
\label{webportal}
The web portal includes additional details on tips for first time conference attendees and their mentors, tables of societies and scientific conferences, and affinity group organizations. 
  \url{https://sites.google.com/macalester.edu/simplerules/home}. 

\section*{Acknowledgments}
We thank Erik Zornik and Andrew Bray for their early contributions to the conference advice template. This work is supported by the National Science Foundation (DBI-1750981 to AR).


\begin{thebibliography}{10}

\bibitem{olena2020covid}
Olena A.
\newblock {COVID-19} Ushers in the Future of Conferences.
\newblock The Scientist. 2020.

\bibitem{sarabipour2020research}
Sarabipour S.
\newblock Research Culture: Virtual conferences raise standards for
  accessibility and interactions.
\newblock Elife. 2020;9:e62668.

\bibitem{sarabipour2021changing}
Sarabipour S, Khan A, Seah YFS, Mwakilili AD, Mumoki FN, S{\'a}ez PJ, et~al.
\newblock Changing scientific meetings for the better.
\newblock Nature Human Behaviour. 2021; p. 1--5.

\bibitem{ligozat2020ten}
Ligozat AL, N{\'e}v{\'e}ol A, Daly B, Frenoux E.
\newblock Ten simple rules to make your research more sustainable.
\newblock PLoS Computational Biology. 2020;16(9):e1008148.

\bibitem{reardon2017latest}
Reardon S.
\newblock How the latest US travel ban could affect science.
\newblock Nature News. 2017;550(7674):17.

\bibitem{salomon2020future}
Salomon D, Feldman MF.
\newblock The future of conferences, today: Are virtual conferences a viable
  supplement to ``live'' conferences?
\newblock EMBO reports. 2020;21(7):e50883.

\bibitem{rich2020ten}
Rich S, Diaconescu AO, Griffiths JD, Lankarany M.
\newblock Ten simple rules for creating a brand-new virtual academic meeting
  (even amid a pandemic).
\newblock PLOS Computational Biology. 2020;16(12):1--9.
\newblock doi:{10.1371/journal.pcbi.1008485}.

\bibitem{arnal2020ten}
Arnal A, Epifanio I, Gregori P, Mart{\'\i}nez V.
\newblock Ten Simple Rules for organizing a non-real-time web conference.
\newblock PLOS Computational Biology. 2020;16(3):1--13.
\newblock doi:{10.1371/journal.pcbi.1007667}.

\bibitem{gichora2010ten}
Gichora NN, Fatumo SA, Ngara MV, Chelbat N, Ramdayal K, Opap KB, et~al.
\newblock Ten simple rules for organizing a virtual conference-anywhere.
\newblock PLoS Computational Biology. 2010;6(2):e1000650.

\bibitem{lortie2020online}
Lortie CJ.
\newblock Online conferences for better learning.
\newblock Ecology and evolution. 2020;10(22):12442--12449.

\bibitem{mabrouk2009survey}
Mabrouk PA.
\newblock Survey study investigating the significance of conference
  participation to undergraduate research students.
\newblock Journal of Chemical Education. 2009;86(11):1335.

\bibitem{davis2017supporting}
Davis J, Alvarado C.
\newblock Supporting undergraduates to make the most of conferences.
\newblock ACM Inroads. 2017;8(3):32--35.

\bibitem{wright2019can}
Wright HM, Tamer NB.
\newblock Can sending first and second year computing students to technical
  conferences help retention?
\newblock In: Proceedings of the 50th ACM Technical Symposium on Computer
  Science Education; 2019. p. 56--62.

\bibitem{smith2013mentoring}
Smith B.
\newblock Mentoring at-risk students through the hidden curriculum of higher
  education.
\newblock Lexington Books; 2013.

\bibitem{martin2014ten}
Martin JL.
\newblock Ten simple rules to achieve conference speaker gender balance.
\newblock PLoS Computational Biology. 2014;10(11):e1003903.

\bibitem{le2020analysis}
Le TT, Himmelstein DS, Anderson AAH, Gazzara MR, Greene CS.
\newblock Analysis of ISCB honorees and keynotes reveals disparities.
\newblock bioRxiv. 2020;.

\bibitem{shishkova2017gender}
Shishkova E, Kwiecien NW, Hebert AS, Westphall MS, Prenni JE, Coon JJ.
\newblock Gender diversity in a STEM subfield -- analysis of a large scientific
  society and its annual conferences.
\newblock Journal of The American Society for Mass Spectrometry.
  2017;28(12):2523--2531.

\bibitem{tanner2007cultural}
Tanner K, Allen D.
\newblock Cultural competence in the college biology classroom.
\newblock CBE -- Life Sciences Education. 2007;6(4):251--258.

\bibitem{AACU}
{Association of American Colleges \& Universities}. Cultural Competency
  Resources: Teaching to Increase Diversity and Equity in STEM (TIDES); 2021.
\newblock Available from: \url{https://www.aacu.org/tides/cultural-competency}.

\bibitem{singleton2020open}
Singleton KS, Tesfaye R, Dominguez EN, Dukes AJ.
\newblock An open letter to past, current and future mentors of Black
  neuroscientists.
\newblock Nature Reviews Neuroscience. 2020; p. 1--2.

\bibitem{tulshyan2021stop}
Tulshyan R, Burey JA.
\newblock Stop Telling Women They Have Imposter Syndrome.
\newblock Harvard Business Review. 2021;.

\bibitem{jackson2019smiling}
Jackson L.
\newblock The smiling philosopher: Emotional labor, gender, and harassment in
  conference spaces.
\newblock Educational Philosophy and Theory. 2019;.

\bibitem{favaro2016your}
Favaro B, Oester S, Cigliano JA, Cornick LA, Hind EJ, Parsons E, et~al.
\newblock Your science conference should have a code of conduct.
\newblock Frontiers in Marine Science. 2016;3:103.

\end{thebibliography}

\bibliographystyle{plainnat}

\end{document}